\documentclass[pdflatex,sn-chicago]{sn-jnl}

\usepackage{graphicx}%
\usepackage{multirow}%
\usepackage{amsmath,amssymb,amsfonts}%
\usepackage{amsthm}%
\usepackage{mathrsfs}%
\usepackage[title]{appendix}%
\usepackage{xcolor}%
\usepackage{textcomp}%
\usepackage{manyfoot}%
\usepackage{booktabs}%
\usepackage{algorithm}%
\usepackage{algorithmicx}%
\usepackage{algpseudocode}%
\usepackage{listings}
\usepackage{booktabs}
\usepackage{makecell}
\usepackage{caption}
\usepackage{geometry}
\usepackage{braket}
\theoremstyle{thmstyleone}%
\theoremstyle{thmstyletwo}%

\theoremstyle{thmstylethree}%

\begin{document}

\title[Bell Meets General Philosophers of Science]{Bell Meets General Philosophers of Science : Reassessing Measurement Independence}


\author*[1]{\fnm{Yuichiro} \sur{Kitajima}}\email{kitajima.yuichirou@nihon-u.ac.jp}\equalcont{ORCID: \texttt{https://orcid.org/0000-0001-8804-4814}}






\abstract{Bell's inequality is derived from three assumptions: measurement independence, outcome independence, and parameter independence. Among these, measurement independence, often taken for granted, holds that hidden variables are statistically uncorrelated with measurement settings. Under this assumption, the violation of Bell's inequality implies that either outcome independence or parameter independence fails to hold, meaning that local hidden variables do not exist. In this paper, we refer to this interpretive stance as the nonfactorizable position. In contrast, superdeterminism represents the view that measurement independence does not hold. Despite its foundational role, this assumption has received relatively little philosophical scrutiny. This paper offers a philosophical reassessment of measurement independence through three major frameworks in the philosophy of science: de Regt's contextual theory of scientific understanding, Kuhn's criteria for theory choice, and Lakatos's methodology of scientific research programmes. Using these lenses, we evaluate the two major responses to the violation of Bell's inequality, the nonfactorizable position and superdeterminism, and argue that the nonfactorizable position currently fares better across all three criteria. Beyond this binary, we introduce a spectrum of intermediate positions that allow for partial violations of measurement independence, modeled via mutual information. These positions modify the ``positive heuristic'' of superdeterminism, a crucial component in Lakatos's definition of research programmes, offering avenues for progressive research. This analysis reframes the debate surrounding Bell's inequality and illustrates how methodological tools can effectively guide theory evaluation in physics.

}




\maketitle

\section{Introduction}
\label{introduction-section}

Quantum theory challenges classical intuitions by allowing a measurement in one region to influence a distant outcome, a phenomenon called nonlocality. Bell's inequalities, such as the CHSH version \citep{clauser1969proposed}, test this effect, and \citet{aspect1982experimental_test} provided decisive evidence that violating them indicates nonlocality. In experiments testing the CHSH inequality, measurements are conducted in two spatially separated regions. The inequality is derived under three key assumptions \citep{myrvold2024bell}:

\begin{description} 
\item[Outcome Independence:] Given hidden variables, the outcome of one measurement does not influence the probability of the other measurement outcome. 
\item[Parameter Independence:] Given hidden variables, the setting of one measurement device does not influence the probability of the outcome at the other device. 
\item[Measurement Independence:] The settings of the measurement devices are statistically uncorrelated with the hidden variables. 
\end{description}

While the first two are associated with locality, measurement independence is often taken for granted and less scrutinized. Measurement independence is the idea that the settings of measurement devices are statistically independent of the underlying hidden variables. Terminological variation exists: some authors use ``statistical independence'' \citep{hossenfelder2020rethinking} or ``setting independence'' \citep{muller2023condition}, but we adopt the term ``measurement independence'' consistently.


Following \citet{myrvold2024bell}, we adopt the term `factorizability' to denote the joint assumption of parameter independence and outcome independence. The `nonfactorizable position' thus refers to interpretations that deny either (or both) of these assumptions while preserving measurement independence.

If the CHSH inequality is violated, at least one of the three assumptions must fail. The nonfactorizable position assumes measurement independence and attributes the violation to a failure of either outcome or parameter independence. In contrast, superdeterminism rejects measurement independence while preserving the other two assumptions. It claims that measurement settings are determined by past hidden variables, possibly originating from the Big Bang \citep{brans1988bell, hossenfelder2020rethinking, nikolaev2023aspects}. Thus, superdeterminism interprets measurement dependence not as epistemological, but as ontological. Table \ref{tab:comparison1} below captures this contrast.



\begin{table}[htbp]
    \centering
    \renewcommand{\arraystretch}{1.3} 
    \small 
    \begin{tabular}{lcc}
        \toprule
        & \textbf{\large Nonfactorizable} & \textbf{\large Superdeterminism} \\
        & \textbf{\large Position} &  \\
        \midrule
        \textit{Measurement Independence} & Holds & Does not hold \\
        \midrule
        \textit{Measurement Device Settings} & \makecell[c]{Randomly\\chosen} & \makecell[c]{Predetermined\\at the Big Bang} \\
        \bottomrule
    \end{tabular}
    \caption{Comparison between the Nonfactorizable Position and Superdeterminism}
    \label{tab:comparison1}
\end{table}

This prevailing dichotomy between nonfactorizable and superdeterministic interpretations may obscure a richer spectrum of intermediate possibilities. This paper proposes intermediate frameworks that permit partial violations of measurement independence, thereby offering a conceptual bridge between these two extremes.



While measurement independence is often treated as a technical assumption in discussions of the CHSH inequalities, this paper emphasizes that its philosophical interpretation plays a decisive role in evaluating the foundations of quantum theory.
Specifically, whether measurement independence holds or not can be interpreted from at least two distinct perspectives:
\begin{itemize}
\item Ontologically, as a claim about the causal structure of the universe (e.g., whether measurement settings are influenced by hidden variables), or
\item Epistemologically, as a statement about our knowledge or ignorance of such influences.
\end{itemize}

This distinction is central to the disagreement between the nonfactorizable position and superdeterminism. It also serves as a key axis for analysis throughout this paper.

Superdeterminism interprets measurement dependence as an ontological feature, whereas the nonfactorizable position regards measurement independence as epistemologically significant. Later sections will distinguish between ontological and epistemological interpretations of measurement dependence, a distinction crucial to understanding intermediate positions between the nonfactorizable position and superdeterminism.


Although there are various criticisms of superdeterminism, such as those related to free will\footnote{One common criticism of superdeterminism is that it denies the existence of free will since superdeterminism assumes that measurement settings are determined by hidden variables originating at the Big Bang. \citet[Section 4.1]{hossenfelder2020rethinking}, however, counters that determinism and free will can be compatible, depending on how free will is defined.}, conspiratorial initial conditions\footnote{One common criticism of superdeterminism is that it has a conspiratorial character since superdeterminism assumes that measurement settings are determined by hidden variables originating at the Big Bang. However, \citet[Section 3]{andreoletti2022superdeterminism} argues that even if superdeterminism appears conspiratorial, it remains logically consistent.}, and scientific methodology\footnote{\citet[p. 168]{shimony1993searchII} and \citet[p. 192]{chen2021bell} warn that if measurement settings truly cannot be chosen at random, as superdeterminism suggests, then the validity of experiments in other scientific fields would be undermined. For example, in a study on the effects of tobacco smoke on mice, researchers typically assign mice to experimental and control groups at random. Without such random assignment, these experiments would lose their integrity. \citet[Section 4.2]{baas2023does} raises a similar concern.

In response, superdeterminists question whether this assumption of random measurement settings is justifiable in the context of quantum physics. For instance, \citet[p. 8]{hossenfelder2020rethinking} argues that measurement independence cannot be taken for granted in quantum systems. Furthermore, \citet[Section 5]{andreoletti2022superdeterminism} suggests that even if correlations between hidden variables and measurement settings exist, they might be practically undetectable.}, this paper does not address these discussions. Instead, the focus is on comparing interpretative strategies through philosophical frameworks. Accordingly, in addition to \citet{kuhn1977essential}'s theory choice criteria and \citet{lakatos1978methodology}'s methodology of research programmes, this paper also considers which position is more intelligible from the perspective of scientific understanding, as developed by \citet{de2017understanding}.

\citet{de2017understanding}'s is not the only approach to scientific understanding. For example, \citet{khalifa2017understanding, khalifa2022onwards, khalifa2022should} offers an alternative view, defining ``understanding as knowledge of explanations'' \citep[p. 17]{de2022can}. This contrasts with de Regt's view, which emphasizes not only knowledge but also the skills needed to achieve understanding. However, through dialogue with \citet{de2022can, de2022frenemies}, \citet[p. 57]{khalifa2022onwards} has come to acknowledge that ``explanatory knowledge requires explanatory abilities'', indicating a convergence of views. Therefore, this paper adopts de Regt's approach as its framework.


This paper applies three influential frameworks in the philosophy of science: de Regt's theory of scientific understanding, Kuhn's theory choice criteria, and Lakatos's methodology of research programmes, to evaluate competing interpretations of Bell inequality violations. These frameworks are deliberately chosen because they represent complementary epistemic perspectives. Taken together, they allow for a multi-dimensional assessment of theory appraisal, how theories are understood (de Regt), chosen (Kuhn), and developed (Lakatos), which is particularly valuable in foundational debates where empirical arbiters are limited\footnote{Superdeterminism has yet to offer a comprehensive account of how hidden variables, presumably originating at the Big Bang, determine the settings of measurement devices. Some might argue that such an incomplete theory cannot be evaluated using the frameworks of de Regt, Kuhn, and Lakatos. However, even incomplete theories can be analyzed using these frameworks. For example, \citet[Section 6.2]{de2017understanding} discusses the early kinetic theory of gases, which did not yet match theoretical and experimental values; \citet[p. 323]{kuhn1977essential} addresses Copernicus's astronomical theory in its early, imprecise and undeveloped form; and \citet[p. 53]{lakatos1978methodology} examines Prout's theory of atomic weights, which faced numerous anomalies. In other words, these frameworks are intended to apply even to incomplete theories. Therefore, even if superdeterminism is not yet fully developed, it is still analyzable within these frameworks.}. This choice is motivated by the paper's focus on methodological evaluation rather than metaphysical interpretation, especially under conditions of empirical underdetermination.

The nonfactorizable position emerges as the most compelling. This conclusion is supported when viewed through the lenses of understanding, theory choice, and research progressiveness. However, the aim is not to claim that these criteria are absolute. Rather, the value of introducing these perspectives lies in the way they open up the possibility of examining measurement independence not simply as a binary opposition between the nonfactorizable position and superdeterminism, but also from an intermediate standpoint that differs from both.

Such intermediate positions emerge through modifications of the positive heuristic of superdeterminism. The positive heuristic is one of the core components of a research programme, as defined by Lakatos. A research programme that incorporates such a revision has the potential to become progressive \citep[p. 51]{lakatos1978methodology}.

The structure of this paper is as follows. Section \ref{experiment} provides an overview of experimental tests of the CHSH inequality from the perspective of measurement independence. 
Sections \ref{de Regt-section}, \ref{kuhn-comparison-subsection}, and \ref{Lakatos-section} compare and examine the nonfactorizable position and superdeterminism. 

Section \ref{de Regt-section} analyzes measurement independence through de Regt's account of scientific understanding; Section \ref{kuhn-comparison-subsection} examines it using Kuhn's theory choice criteria; and Section \ref{Lakatos-section} approaches it through Lakatos's methodology of research programmes. Based on these frameworks, the analysis suggests that the nonfactorizable position currently appears more promising. 

These three criteria do not lead to the conclusion that the nonfactorizable position is absolutely correct. In Section \ref{cautions-section}, we examine the limitations of these three criteria. However, examining the factorizable position and superdeterminism through these lenses reveals something important: the existence of intermediate positions between the two. As discussed in Section \ref{intermediate-section}, such intermediate positions emerge through modifications to the positive heuristic of superdeterminism, where the positive heuristic is one of the key components of a research programme as defined by Lakatos.

Modifying the positive heuristic alone leads only to qualitative discussions. In Section \ref{correlation-subsubsection}, we build on the work of \citet{barrett2011much} and \citet{hall2011relaxed, hall2016significance} by using the quantity $C_{MD}$, which quantifies measurement dependence through mutual information. While they have introduced $C_{MD}$ as a quantitative tool within quantum foundations, their work does not engage with the philosophical frameworks of theory appraisal. This paper is the first to integrate $C_{MD}$ into a structured philosophical comparison between rival interpretations.

We examines whether $C_{MD}$ should be interpreted ontologically or epistemologically and situates it accordingly within a philosophical context (Section \ref{ontological-epistemological-subsection}). Positions that interpret $C_{MD}$ epistemologically provide an epistemically modest yet conceptually rich alternative that bridges foundational philosophy with quantum information theory. This analysis not only reframes existing debates about Bell's inequality but also illustrates how philosophical tools can deepen our understanding of theory appraisal in physics.



\section{Experimental Tests of the CHSH Inequality}
\label{experiment}

Numerous experiments have tested violations of the CHSH inequality, typically under the assumption that measurement independence holds. This section briefly summarizes those efforts and highlights the persistent philosophical ambiguity about whether measurement settings are truly uncorrelated with hidden variables.

\citet{aspect1982experimental_test} provided early evidence for CHSH violations using entangled photons \citep{brunner2014bell, scarani2019bell}. Although their experiment improved on earlier designs by varying measurement settings during photon flight, the settings themselves were still determined in advance, leaving room for hidden-variable correlations. Later experiments, e.g., by \citet{weihs1998violation} and \citet{scheidl2010violation}, employed fast random number generators to ensure that settings were chosen during flight. These methods aimed to secure statistical independence, between measurement choices and any local hidden variables.

Nevertheless, certain logical loopholes persist. As  \citet[p. 61]{bell2004speakable} noted, even seemingly random choices may be correlated with hidden variables if those variables exist within the shared past light cones of the settings and the particles. To address this, cosmic photon-based experiments used starlight from quasars billions of years away as randomizers, pushing the possible correlations further back in time \citep{rauch2018cosmic}. While many researchers find such experimental designs persuasive, they cannot strictly rule out the possibility of superdeterminism. This interpretation posits that measurement settings were determined by hidden variables originating at the Big Bang  \citep[Section 2.2]{nikolaev2023aspects}.

Several theorists have attempted to model local violations of the CHSH inequality by rejecting measurement independence. For instance, \citet{t2016cellular} proposed a deterministic cellular automaton underlying quantum states, while \citet{donadi2022toy} developed a toy model incorporating hidden-variable correlations. However, \citet[p. 3]{t2016cellular} and \citet[p. 022212-1, p. 022212-10]{donadi2022toy} acknowledge that their models are highly speculative, lack empirical distinctiveness, and do not yield new predictions. Similarly, \citet{ciepielewski2023superdeterministic}  proposed a model that is fully local and consistent with quantum predictions, but at the cost of extreme complexity and empirical indistinguishability from standard quantum mechanics.

These experiments and models demonstrate that violating measurement independence remains a logically viable explanation for Bell inequality violations. Yet such an approach continues to face challenges of testability, conceptual simplicity, and empirical distinctiveness, setting the stage for the philosophical analyses that follow.

\section{de Regt's Framework: Evaluating Scientific Understanding}
\label{de Regt-section}

This section aims to evaluate the philosophical viability of the assumption of measurement independence, particularly through the lens of de Regt's theory of scientific understanding. Building on the empirical context provided in Section \ref{experiment}, we ask: which interpretation, the nonfactorizable position or superdeterminism, better supports scientific intelligibility?



\subsection{de Regt's Account of Scientific Understanding}

In the philosophy of science, the question of what constitutes a scientific explanation has been extensively discussed by many scholars, with classic models such as Hempel's Deductive-Nomological (D-N) model representing the logical positivist tradition. In contrast to scientific explanation, the concept of scientific understanding has traditionally received less philosophical analysis, particularly regarding its independent epistemic value. \citet{trout2007psychology}, for example, argues that understanding lacks independent epistemic significance. In his view, it is merely a byproduct of scientific explanation.

In opposition to this view, \citet{de2017understanding} argues that scientific understanding has epistemic significance and proposes a contextual approach. He distinguishes among three components: the sense of understanding, understanding of phenomena, and understanding of theory. Of these, the ``sense of understanding'' is regarded as subjective and unreliable. The more crucial aspects, according to de Regt, are the latter two.

According to \citet{de2017understanding}, a phenomenon is understood when it is explained using an intelligible theory. Such a theory allows scientists to qualitatively understand its implications without requiring exact calculations. This concept forms the basis for two core criteria introduced by de Regt: the Criterion for the Understanding of Phenomena (CUP), which evaluates whether a theory allows scientists to explain observable phenomena; and the Criterion of Intelligibility of Theory (CIT), which assesses whether scientists can use the theory without requiring exact calculations:

\begin{description}
\item[CUP:] 
A phenomenon P is understood scientifically if and only if there is an explanation of P that is based on an intelligible theory T and conforms to the basic epistemic values of empirical adequacy and internal consistency. \citep[p. 92]{de2017understanding}
\end{description}

In this context, ``intelligible'' refers to the following:

\begin{description}
\item[CIT:] A scientific theory T (in one or more of its representations) is intelligible for scientists (in context C) if they can recognize qualitatively characteristic consequences of T without performing exact calculations.
\cite[p. 102]{de2017understanding}
\end{description}

In the cited passage, the phrase ``in one or more of its representations'' means the following: in order to establish a connection between a phenomenon and a theory, appropriate idealizations and approximations must be made. The model in which the theory is represented may vary depending on the skills of the scientist and the features they consider important \cite[p. 36]{de2017understanding}. The Criterion of Intelligibility (CIT) must be satisfied in one or more of its representations. That is the intended meaning of ``in one or more of its representations.''

As an example of qualitative reasoning, \citet[p. 104]{de2017understanding} cites the case of Boltzmann. By assuming the existence of discrete particles, one can qualitatively reason that an increase in temperature leads to an increase in pressure. In such cases, the theory is considered intelligible. This is because it enables qualitative predictions that follow from its core assumptions.


From the perspective of de Regt's framework, we examine which of the nonfactorizable position or superdeterminism is more intelligible in Section \ref{deRegt-comparison-subsection}.

\subsection{Quantum Teleportation}
\label{teleportation-section}

To evaluate which view better satisfies de Regt's intelligibility criterion in Section \ref{deRegt-comparison-subsection}, this section provides a brief overview of quantum teleportation as a paradigmatic manifestation of quantum nonlocality. This will serve as a benchmark against which the intelligibility of each interpretation can be assessed.


The violation of the CHSH inequality has been extensively discussed in philosophical literature. For example, \citet{howard1985einstein, howard1989holism} and \citet{teller1989relativity} explored its philosophical implications by arguing that while parameter independence holds, outcome independence does not. In contrast to such philosophical analyses, a different line of research treats the violation of the CHSH inequality as a physical resource. The interpretation of the CHSH inequality violations as physical resources has led to developments in quantum information theory, including quantum teleportation \citep{bennett1993teleporting}. Quantum teleportation is not merely a theoretical prediction; it has also been experimentally demonstrated by \citet{bouwmeester1997experimental}. Subsequent experiments have also demonstrated quantum teleportation using continuous variables \citep{pirandola2015advances}.

By using previously shared the EPR-Bohm state, which is a type of state that violates the CHSH inequality, quantum teleportation enables Alice to send an unknown quantum state to Bob with only classical communication. This process does not transmit matter or energy but transfers the `information content' of the quantum state, as \citet{timpson2006grammar} provided a detailed discussion of how information is interpreted in the context of quantum teleportation.

Suppose Alice and Bob are located far apart in space. Quantum teleportation enables a form of information transfer that is impossible with classical communication alone. Let's concretely examine the procedure of quantum teleportation and clarify how it differs from the CHSH inequality framework.

Let $\ket{0}$ and $\ket{1}$ be unit vectors in a two-dimensional Hilbert space, defined as $\ket{0}=\begin{pmatrix} 1 \\ 0 \end{pmatrix}$ and $\ket{1}=\begin{pmatrix} 0 \\ 1 \end{pmatrix}$. These represent the spin-up and spin-down states along the z-axis, respectively. Alice holds a state 
\begin{equation}
\label{qubit}
\ket{\psi}=a\ket{0}+b\ket{1}.
\end{equation}
Additionally, Alice and Bob share the entangled state
\begin{equation}
\label{EPR-state}
\ket{\phi_{00}}=\frac{1}{\sqrt{2}}\ket{0}\ket{0}+\frac{1}{\sqrt{2}}\ket{1}\ket{1}.
\end{equation}
Thus, the total combined state is:
\begin{equation}
\begin{split}
\ket{\psi}\ket{\phi_{00}} = \frac{1}{2} &\Bigg( \frac{\ket{0}\ket{0}+\ket{1}\ket{1}}{\sqrt{2}}(a\ket{0}+b\ket{1}) \\
&+\frac{\ket{0}\ket{0}-\ket{1}\ket{1}}{\sqrt{2}}(a\ket{0}-b\ket{1}) \\
&+\frac{\ket{0}\ket{1}+\ket{1}\ket{0}}{\sqrt{2}}(a\ket{1}+b\ket{0}) \\
&+\frac{\ket{0}\ket{1}-\ket{1}\ket{0}}{\sqrt{2}}(a\ket{1}-b\ket{0}) \Bigg) .
\end{split}
\end{equation}
Alice performs a measurement known as a Bell state measurement, which identifies which of the following states the system is in:
\begin{equation}
\label{bell-state-measurement}
\Bigg\{ \frac{\ket{0}\ket{0}+\ket{1}\ket{1}}{\sqrt{2}}, \frac{\ket{0}\ket{0}-\ket{1}\ket{1}}{\sqrt{2}}, \frac{\ket{0}\ket{1}+\ket{1}\ket{0}}{\sqrt{2}}, \frac{\ket{0}\ket{1}-\ket{1}\ket{0}}{\sqrt{2}} \Bigg\}.
\end{equation}
For example, suppose Alice obtains the state $(\ket{0}\ket{1}+\ket{1}\ket{0})/\sqrt{2}$ as the result of her measurement. She communicates the result to Bob, who then applies the appropriate unitary transformation. Specifically, he applies $\sigma_x=\begin{pmatrix} 0 & 1 \\ 1 & 0 \end{pmatrix}$ to his state. As a result:
\begin{equation}
\sigma_x(a\ket{1}+b\ket{0}) =a\ket{0}+b\ket{1},
\end{equation}
which reconstructs Alice's original state. This protocol is called quantum teleportation.

On the other hand, in the context of the CHSH inequality violations, Alice chooses one of two possible measurements, for example, the following, and performs the measurement.:
\begin{equation}
\label{bell-measurement-1}
\begin{pmatrix} 1 & 0 \\ 0 & -1 \end{pmatrix} \otimes I
\end{equation}
or
\begin{equation}
\label{bell-measurement-2}
\left( \cos \frac{\pi}{4} \begin{pmatrix} 1 & 0 \\ 0 & -1 \end{pmatrix} + \sin \frac{\pi}{4} \begin{pmatrix} 0 & 1 \\ 1 & 0 \end{pmatrix} \right) \otimes I.
\end{equation}

These matrices do not commute, so they cannot be observed simultaneously. The nonfactorizable position assumes that the choice between these two measurements is made randomly, while superdeterminism posits that hidden variables determine which measurement is made.

In contrast, in the case of quantum teleportation, only the measurement represented by Equation (\ref{bell-state-measurement}) is performed. There is no procedure involving a choice between, for example, the measurement in Equation (\ref{bell-measurement-1}) and that in Equation (\ref{bell-measurement-2}). This marks a key difference between quantum teleportation and the CHSH inequality scenario.

\subsection{Assessing the nonfactorizable Position and Superdeterminism from the Perspective of Intelligibility}
\label{deRegt-comparison-subsection}

This section uses \citet{de2017understanding}'s framework to compare the nonfactorizable position and superdeterminism, focusing on their qualitative explanatory power and intelligibility.

While de Regt's contextual approach offers useful tools for evaluating intelligibility, it is not without its critics. For example, scholars like \citet{trout2007psychology} argue that understanding has no independent epistemic role. This paper aligns with de Regt's position but acknowledges that different criteria may lead to different conclusions. The reason for adopting multiple criteria is precisely to recognize this point.

Suppose Alice and Bob, who are located at spatially separated positions, share the state $\ket{\phi_{00}}$ (Equation (\ref{EPR-state})). As mentioned above, quantum teleportation involves only one type of measurement, as given in Equation (\ref{bell-state-measurement}). Therefore, unlike in the CHSH inequality scenario, it is not possible to explain the process in terms of hidden variables that determine the measurement choice. In other words, quantum teleportation cannot be explained through a violation of measurement independence.

In quantum teleportation, the state $\ket{\phi_{00}}$, represented by Equation (\ref{EPR-state}), is shared between Alice and Bob. This state violates the CHSH inequality. According to the nonfactorizable position, a state that violates the CHSH inequality is nonclassical in the sense that either outcome independence or parameter independence fails. This implies that Alice and Bob share a nonclassical state, which enables them to perform quantum teleportation, something that is impossible under classical physics.

In contrast, according to superdeterminism both outcome and parameter independence hold, even in states that violate the CHSH inequality. Thus $\ket{\phi_{00}}$ is a classical state. Therefore, it does not permit the same kind of qualitative reasoning as the nonfactorizable position does.

Consequently, the nonfactorizable position is considered intelligible, whereas superdeterminism is not.

\section{Assessment Based on Kuhn's Criteria for Theory Choice}
\label{kuhn-comparison-subsection}


\citet[pp. 321--322]{kuhn1977essential} proposes five criteria for theory choice: accuracy, consistency, scope, simplicity, and fruitfulness. \citet[p. 37]{de2017understanding} notes that Kuhn's criteria for theory choice are also relevant to evaluating a theory's intelligibility. While these are not meant as strict rules, they serve as useful heuristics for evaluating competing interpretations of quantum theory.

Both the nonfactorizable position and superdeterminism perform equally well in terms of accuracy and scope, as each is designed to reproduce quantum predictions, including CHSH violations. The key differences emerge in the other three criteria.

On simplicity, the picture is mixed. The nonfactorizable position requires no additional assumptions beyond standard quantum theory. Superdeterminist models, by contrast, often invoke complex hidden-variable mechanisms as discussed in Section \ref{experiment}. However, some argue that superdeterminism is simpler at a deeper metaphysical level, since it postulates a fully deterministic universe. Because of this ambiguity, simplicity by itself cannot determine which position is preferable.

Regarding consistency, superdeterminism appears problematic. \citet[Section 6]{lewis2006conspiracy} argues that superdeterminism is inconsistent. Consider an experiment testing the violation of the CHSH inequality. In this experiment, the settings of the measuring devices could be determined either by a human or by a lottery machine. These are physically distinct processes. Nevertheless, in both cases, the hidden variables would need to be correlated with the device settings. This means the correlation depends not on the physical nature of the process. Instead, it depends solely on whether the process is part of a CHSH experiment. \citet[Section 6]{lewis2006conspiracy} argues that superdeterminism based on such non-physical distinctions is inconsistent. In contrast, the nonfactorizable position avoids such problems because the hidden variables are not correlated with the measurement settings.


Finally, in terms of fruitfulness, the nonfactorizable position has been notably productive. It has inspired foundational advances in quantum information science such as quantum teleportation as discussed in Section \ref{deRegt-comparison-subsection}. Superdeterminism, despite being logically consistent, has yet to yield new predictions or practical applications.

\citet[p. 40]{de2017understanding} notes that intelligibility functions as a contextual measure of fruitfulness. This means that what counts as fruitful can vary depending on the scientific context. In one context, a theory may be highly fruitful, while in another, it may have little explanatory or predictive value.

In the context of the nonfactorizable position, intelligibility has proven to be highly fruitful. Because the theory is intelligible, it has led to concrete developments such as quantum information science. These applications emerged precisely because the underlying nonclassical features of quantum theory could be grasped and utilized, both conceptually and practically.

By contrast, superdeterminism lacks this kind of fruitfulness. It does not provide intelligible, qualitative predictions that could lead to new phenomena or technological developments. As a result, it has seen little theoretical progress and few practical applications. In this sense, superdeterminism cannot be considered fruitful.

Table \ref{tab:Kuhn-comparison} summarizes the comparison. From the perspectives of fruitfulness and consistency, the nonfactorizable position appears more promising than superdeterminism.

\begin{table}[htbp]
    \centering
    \renewcommand{\arraystretch}{1.3}
    \small
    \begin{tabular}{lcc}
        \toprule
        & \textbf{\large Nonfactorizable Position} & \textbf{\large Superdeterminism} \\
        \midrule
         \textit{Fruitfulness} & \makecell[c]{Contributed to the development\\of quantum information theory} & \makecell[c]{No established\\practical applications} \\
         \midrule
        \textit{Consistency} & -- & \makecell[c]{Deciding which measuring devices \\ are correlated with hidden variables \\ based on distinctions that \\ have no physical definition} \\
        \midrule
        \textit{Simplicity (1)} & No additional theory needed & \makecell[c]{Complex additional\\theory required} \\
        \midrule
        \textit{Simplicity (2)} & -- & \makecell[c]{Deterministic worldview\\may be seen as simpler\\in some sense} \\
        \midrule
         \textit{Accuracy} & Matches experimental results & Matches experimental results \\
         \midrule
         \textit{Scope} & Same scope & Same scope \\
        \bottomrule
    \end{tabular}
    \caption{Comparison Based on Kuhn's Criteria}
    \label{tab:Kuhn-comparison}
\end{table}

\section{Evaluation from the Perspective of Lakatos's Research Programme}
\label{Lakatos-section}

Lakatos proposed criteria for determining whether a research programme is progressive (Section \ref{lakatos-subsection}). In Section \ref{lakatos-comparison-subsection}, based on the preceding discussion, we conclude that the nonfactorizable position qualifies as a progressive research programme. 

\subsection{Lakatos's Research Programme}
\label{lakatos-subsection}

Lakatos reformulated the scientific enterprise in terms of research programmes. A research programme consists of several components: 

\begin{itemize}
\item A hard core, made up of fundamental assumptions that are not to be abandoned.
\item A protective belt, comprising auxiliary hypotheses that can be adjusted in response to empirical challenges.
\item A negative heuristic, which defines areas that should not be questioned.
\item A positive heuristic, which offers guidance for productive lines of inquiry.
\end{itemize}

For example, in the research programme of Newton's gravitational theory, the negative heuristic consists in not rejecting ``Newton's three laws of dynamics and his law of gravitation'' \citep[p. 48]{lakatos1978methodology}. The positive heuristic, on the other hand, is encapsulated in the idea that ``the planets are essentially gravitating spinning-tops of roughly spherical shape'' \citep[p. 51]{lakatos1978methodology}. However, these heuristics are flexible and may be subject to change \citep[p. 51]{lakatos1978methodology}.


Lakatos also introduced the distinction between progressive and degenerative research programmes. A theory is considered theoretically progressive if it has ``some excess empirical content over its predecessor, that is, if it predicts some novel, hitherto unexpected fact'' \citep[p. 33]{lakatos1978methodology}. It is empirically progressive if ``some of this excess empirical content is also corroborated, that is, if each new theory leads us to the actual discovery of some new fact'' \citep[p. 34]{lakatos1978methodology}. A theory that is both theoretically and empirically progressive is regarded as progressive overall; otherwise, it is considered degenerative.

Lakatos further states that when a theory comes into conflict with experimental results, ``if we succeed in replacing some ingredient is a `progressive' way (that is, the replacement has more corroborated empirical content than the original), we call it `falsified' '' \citep[p. 41]{lakatos1978methodology}.

Let us once again take Newton's gravitational theory as an example. When anomalies were observed in the orbit of Uranus, the research programme did not revise its hard core, Newton's three laws of dynamics and his law of gravitation, but instead introduced a new auxiliary hypothesis: the existence of an unknown planet. This hypothesis (the prediction of Neptune) was later confirmed by observation. This is a textbook case of a progressive research programme.

In contrast, when modifications are made solely to preserve a theory without leading to new testable predictions, the programme is considered ad hoc and degenerative.

\subsection{Assessment of the Nonfactorizable Position and Superdeterminism from the Perspective of Research Programmes}
\label{lakatos-comparison-subsection}

Lakatos's framework evaluates scientific progress through research programmes. These are organized around a stable `hard core' and a flexible `protective belt.' A programme is deemed progressive if it leads to novel, corroborated predictions. If it merely accommodates existing data without generating new insights, it is considered degenerative.

In this section, the analysis of research programmes is grounded in the historical context surrounding key developments: the emergence of Bell's inequalities in the late 1960s, their experimental verification in the 1980s, the birth of quantum information theory in the 1990s, and the further experimental tests of Bell's inequalities and advances in quantum information theory during the 2000s.

Applying this to the nonfactorizable position, we see a progressive trajectory. Its hard core is the assumption of measurement independence, with violations of the CHSH inequality explained by rejecting outcome or parameter independence. The positive heuristic of the nonfactorizable position is the idea that states violating the CHSH inequality possess nonclassical properties. Motivated by the desire to actively explore these nonclassical features and to test them experimentally, both theoretical and empirical progress has been made, including quantum teleportation\footnote{Quantum teleportation was initially proposed as a protocol in which Alice and Bob share a pure state, such as the one represented by Equation (\ref{EPR-state}). Later, theoretical research also advanced on cases where Alice and Bob share mixed states, known as Werner states \citep{clifton2001nonlocality}. In this way, theoretical research on quantum teleportation has continued to progress.}, quantum cryptography, and novel experimental designs using cosmic sources. Such developments have fulfilled Lakatos's criteria for both theoretical and empirical progress

In contrast, the superdeterminist programme maintains outcome and parameter independence while rejecting measurement independence. The positive heuristic of superdeterminism is the idea that CHSH violations attributed to correlations between measurement settings and hidden variables dating back to the Big Bang\footnote{\citet[p. 51]{lakatos1978methodology} states that a positive heuristic can be revised, and such revisions may transform a degenerating research programme into a progressive one. In Section \ref{classification-subsection}, we explore research programmes that revise the positive heuristic of superdeterminism. These programmes have the potential to become progressive.}. This explanation is logically coherent but has yielded no testable predictions and lacks practical applicability. Models proposed under this framework, such as those by \citet{t2016cellular} or \citet{donadi2022toy}, are often acknowledged by their own authors to be speculative and empirically indistinct from standard quantum theory (Section \ref{experiment}).




\begin{table}[htbp]
    \centering
    \renewcommand{\arraystretch}{1.3}
    \small
    \begin{tabular}{lcc}
        \toprule
        & \textbf{\large Nonfactorizable Position} & \textbf{\large Superdeterminism} \\
        \midrule
        \textit{Hard Core} &  \makecell[c]{Measurement Independence} & \makecell[c]{Outcome Independence\\Parameter Independence} \\
        \midrule
        \textit{Protective Belt} & \makecell[c]{Outcome Independence\\Parameter Independence} & \makecell[c]{Measurement Independence} \\
        \midrule
        \makecell[c]{\textit{Negative}\\ \textit{Heuristic}} & \makecell[c]{Does not deny \\ measurement independence}& \makecell[c]{Does not deny \\ outcome independence and \\ parameter independence}\\
        \midrule
        \makecell[c]{\textit{Positive}\\ \textit{Heuristic}} &  \makecell[c]{States violating \\ the CHSH inequality possess \\ nonclassical properties} & \makecell[c]{Measurement settings\\determined by hidden\\variables at the Big Bang} \\
        \bottomrule
    \end{tabular}
    \caption{Comparison of the Nonfactorizable Position and Superdeterminism}
    \label{tab:comparison7}
\end{table}

The differences between the nonfactorizable position and superdeterminism can be summarized in Table~\ref{tab:comparison7}. From the perspective of Lakatos's positive heuristic, the nonfactorizable interpretation not only provides explanatory power for past quantum phenomena but also functions as a forward-looking guide for research directions. 

One example is the theoretical and experimental development of the so-called `quantum internet', a prospective communication infrastructure that depends on entangled states and teleportation protocols. Such research agendas, fundamentally reliant on the violation of the CHSH inequalities, are consistent with the nonfactorizable position's core assumptions. In contrast, superdeterministic frameworks offer no such generative heuristic.

This comparative analysis highlights the nonfactorizable position as not only methodologically robust but also heuristically fertile, meeting Lakatos's criteria more fully than its rival.

\section{Cautions Regarding the Three Criteria}
\label{cautions-section}

This paper has assessed interpretations of the CHSH inequality violations using three philosophical frameworks: de Regt's theory of understanding, Kuhn's criteria for theory choice, and Lakatos's methodology of research programmes. We concluded that the nonfactorizable position is promising. While these offer valuable insights, none provides a definitive criterion for theory selection.

\citet[Chapter 4]{de2017understanding} emphasizes that scientific understanding is context-dependent, focusing on intelligibility rather than formal derivability. However, this very contextualism raises doubts about whether intelligibility should guide theory choice. A theory may be intelligible within one scientific context but not in another, limiting its general evaluative power. Furthermore, since the claims of superdeterminism are ontological in nature, one might argue that their validity cannot be assessed using de Regt's pragmatic criteria.

\citet[p. 325]{kuhn1977essential} acknowledged that theory choice inevitably involves both objective and subjective elements. He emphasized that no formal algorithm can prescribe how to apply his five criteria in a given case. A scientist prioritizing simplicity of determinism over fruitfulness may reasonably favor superdeterminism, illustrating Kuhn's view that different criteria may yield different choices.

Lakatos offers a more historically structured framework but is similarly cautious. \citet[p. 86]{lakatos1978methodology} explicitly rejects the idea that a single experiment, such as the one conducted by \citet{aspect1982experimental_test}, can falsify an entire research programme, emphasizing instead the cumulative and comparative nature of theory appraisal. 

Still, Lakatos's framework is useful for evaluating the trajectory of research programmes. Since the 1990s, experiments have aimed to close the measurement independence loophole (e.g., via cosmic photons), reinforcing the nonfactorizable position's empirical progressiveness. The rise of quantum information theory (e.g., quantum teleportation) has further extended its theoretical fruitfulness.

That said, Lakatos's research programme framework, by its nature, contains what he himself described as a ``hindsight element'' \cite[p. 70]{lakatos1978methodology}. Superdeterminism could evolve if it were to generate novel, testable prediction, e.g., experimentally detectable correlations between hidden variables and measurement settings. However, such predictions remain elusive.

In sum, while the nonfactorizable position currently fares better under all three frameworks, none of these perspectives offers a final verdict. Kuhn held that the decision regarding which of the five objective criteria to prioritize in theory choice inevitably involves subjective elements. This view is also applicable to the present discussion. One could argue that the decision about which perspective to emphasize, the concept of scientific understanding proposed by \citet[Chapter 4]{de2017understanding}, the criteria for theory choice by \citet[Chapter 13]{kuhn1977essential}, or the methodology of scientific research programmes by \citet[Chapter 1]{lakatos1978methodology}, is similarly shaped by subjective judgment. For example, if one places greater importance on Kuhn's ``simplicity (2)'' criterion, as discussed in Section \ref{kuhn-comparison-subsection}, one might be led to favor superdeterminism. In this sense, the analysis based on these three perspectives cannot be considered definitive.

However, examining the factorizable position and superdeterminism through these lenses, particularly from Lakatos's perspective, reveals something important: the existence of intermediate positions between the two. In the following sections, we examine these intermediate positions in more detail.

\section{Beyond Dichotomy: Intermediate Theories and Quantifying Measurement Dependence}
\label{intermediate-section}

Rather than viewing measurement independence as either entirely valid or wholly violated, this paper explores a spectrum of positions in which partial violations may occur. These ``intermediate positions'' leverage information-theoretic tools, such as mutual information, to quantify the degree of measurement dependence, thereby challenging the prevailing binary between nonlocality and superdeterminism. 

Conceptually, this spectrum allows us to reconceptualize hidden variables not necessarily as cosmic determinants, but as information that may be accessible to local agents, such as eavesdroppers or adversarial systems. This reframing opens new possibilities. It allows us to better understand the limits of physical determinism, while also offering new insights into the foundations of quantum information theory.


These intermediate positions, unlike strong superdeterminism, are not constrained by cosmological determinism and thus offer new pathways for reconciling local hidden variables with quantum predictions. The intermediate positions are summarized in Table \ref{tab:comparison7}. Here, OI stands for outcome independence, PI for parameter independence, and MI for measurement independence.

A common feature of the intermediate positions is that their hard core consists of OI (outcome independence) and PI (parameter independence), while MI (measurement independence) is placed in the protective belt. This structure is the same as in superdeterminism. What distinguishes intermediate positions from superdeterminism is the underlying positive heuristic that guides how hidden variables are interpreted. The positive heuristic of superdeterminism posits that the settings of measuring devices are determined by hidden variables originating at the Big Bang. In contrast, the intermediate positions revise this heuristic. \citet[p. 51]{lakatos1978methodology} states that ``it occasionally happens that when a research programme gets into a degenerating phase, a little revolution or a creative shift in its positive heuristic may push it forward again.''

Intermediate Position 1 (Section \ref{first-intermediata-subsection}) interprets hidden variables as the physical states of the detectors. Accordingly, its positive heuristic holds that the measurement settings are determined by the physical states of the detectors. Like superdeterminism, this view treats hidden variables ontologically (Section \ref{ontological-epistemological-subsection}).

By contrast, Intermediate Positions 2 and 3 (Sections \ref{second-intermediata-subsection} and \ref{third-intermediata-subsection}) adopt an epistemological interpretation of hidden variables (Section \ref{ontological-epistemological-subsection}). In these positions, hidden variables are interpreted as information: in Intermediate Position 2, as information possessed by an eavesdropper; and in Intermediate Position 3, as information held by someone attempting to forge random numbers. The positive heuristic of Intermediate Position 2 assumes that the eavesdropper knows the settings of the measurement devices, while that of Intermediate Position 3 assumes that the forger has knowledge of those settings.

Table \ref{tab:comparison7.1} summarizes the positive heuristics of superdeterminism and Intermediate Positions 1, 2, and 3.



\begin{table}[htbp]
    \centering
    \renewcommand{\arraystretch}{1.3}
    \small
    \begin{tabular}{lccc}
        \toprule
        & \makecell[c]{ \textbf{\large Positive} \\  \textbf{\large Heuristic}} & \makecell[c]{ \textbf{\large Hard} \\ \textbf{\large Core}}& \makecell[c]{\textbf{\large Protective}\\\textbf{\large Belt}} \\
        \midrule
        \textit{Superdeterminism} & \makecell[c]{Measurement settings\\determined by hidden\\variables at the Big Bang} & OI and PI & MI \\
        \midrule
       \makecell[c]{ \textit{Intermediate}\\ \textit{ Position 1}\\ (Section \ref{first-intermediata-subsection})} & \makecell[c]{The measurement settings \\ are determined by \\ the physical states \\ of the detectors} & OI and PI & MI \\
       \midrule
        \makecell[c]{\textit{Intermediate}\\ \textit{ Position 2}\\(Section \ref{second-intermediata-subsection})} & \makecell[c]{The eavesdropper \\ knows the settings of \\ the measurement devices} & OI and PI & MI \\
        \midrule
        \makecell[c]{ \textit{Intermediate}\\ \textit{ Position 3}\\(Section \ref{third-intermediata-subsection})}& \makecell[c]{The forger \\ knows the settings of \\ the measurement devices} & OI and PI & MI \\
        \bottomrule
    \end{tabular}
    \caption{Comparison of Intermediate Positions}
    \label{tab:comparison7.1}
\end{table}

Up to this point, the discussion of measurement independence has been qualitative. In this section, attention has been directed toward research that quantifies the degree to which measurement independence is violated. \citet{barrett2011much}, as well as \citet{hall2011relaxed, hall2016significance}, have investigated this issue by using mutual information as a measure to quantify the violation of measurement independence and examined its relationship with the violation of the CHSH inequalities.

Section \ref{correlation-subsubsection} introduces this approach to quantification. Section \ref{classification-subsection} classifies intermediate positions based on this quantitative framework. Section \ref{ontological-epistemological-subsection} examines two perspectives on the quantity introduced in Section \ref{correlation-subsubsection}: one that interprets it ontologically, and another that interprets it epistemologically. Section \ref{future-prospects-subsection} discusses the significance of this classification and future prospects.

\subsection{Correlation Measure of Dependence}
\label{correlation-subsubsection}

Mutual information, which quantifies how much knowledge of one variable reduces uncertainty about another, is used here to measure the degree to which measurement settings and hidden variables are correlated. It captures how much information one can gain about one variable by knowing the outcome of the other. To make this more accessible, we first illustrate mutual information through a simple coin-flipping example, before applying it to the quantum case.

For example, let $X_A$ be the random variable representing the outcome (heads or tails) of tossing coin A, and $X_B$ be the corresponding random variable for coin B. Let $p(h_A,h_B)$, $p(t_A,t_B)$, $p(h_A,t_B)$, $p(h_A,t_B)$ represent the joint probabilities of obtaining heads-heads, tails-tails, heads-tails, and tails-heads, respectively. Similarly, let $p(h_A)$, $p(t_A)$ be the marginal probabilities of getting heads or tails with coin A, and $p(h_B)$, $p(t_B)$ be the corresponding probabilities for coin B.

Then, the mutual information $I(X_a; X_b)$ between the two variables is given by:
\[ \begin{split}
I(X_a; X_b)
&=p(h_A,h_B)\log_2\frac{p(h_A,h_B)}{p(h_A)p(h_B)}  \\
& \ \ \ +p(t_A,t_B)\log_2\frac{p(t_A,t_B)}{p(t_A)p(t_B)}  \\
& \ \ \ +p(h_A,t_B)\log_2\frac{p(h_A,t_B)}{p(h_A)p(t_B)}  \\
& \ \ \ +p(t_A,h_B)\log_2\frac{p(t_A,h_B)}{p(t_A)p(h_B)}.  \\
\end{split} \]

Suppose that $p(h_A)=p(t_A)=p(h_B)=p(t_B)=0.5$. This indicates that, when viewed individually, coins A and B appear to behave like fair coins.

For example, consider the case where $p(h_A,h_B)=p(t_A,t_B)=0.25$ and $p(h_A,t_B)=p(h_A,t_B)=0.25$. This describes a situation in which both coin A and coin B are fair (i.e., unbiased) and mutually independent. In other words, knowing the outcome of one coin tells us nothing about the outcome of the other. If we interpret $X_b$ as the hidden variable and $X_a$ as the measurement setting, then this exemplifies a situation where measurement independence remains intact. In this case, the mutual information is $I(X_a; X_b)=0$. 

In contrast, consider a case where $p(h_A,h_B)=p(t_A,t_B)=0.5$ and $p(h_A,t_B)=p(h_A,t_B)=0$. Here, the outcomes of coin A and coin B are perfectly correlated: if you know the result of coin B, you know with certainty the result of coin A. That is, $p(h_A|h_B)=p(t_A|t_B)=1$. If we interpret $X_b$ as the hidden variable and $X_a$ as the measurement setting, then this corresponds to a case in which the measurement setting is determined by the hidden variable. In this case, the mutual information is $I(X_a; X_b)=1$.

There are also intermediate cases between the two examples above. For instance, consider the case where $p(h_A,h_B)=p(t_A,t_B)=0.3252$ and $p(h_A,t_B)=p(h_A,t_B)=0.1748$. In this case, $X_A$ and $X_B$ are not independent. Specifically, $p(h_A|h_B)=0.6504 \neq p(h_A)=0.5$ and $p(h_A|t_B)=0.3496 \neq p(h_A)=0.5$, so knowing the result of coin B allows us to make a better-than-chance guess about the outcome of coin A, though not with certainty. The mutual information in this case is approximately $I(X_a; X_b) \approx 0.0663$.

To quantify the extent to which measurement independence is violated, \citet[Equation (28)]{hall2011relaxed} introduced the Correlation Measure of Dependence based on mutual information. Following \citet[Equation (11.10)]{hall2016significance}, we denote this quantity by $C_{MD}$. 

When measurement independence holds, we have $C_{MD}=0$. $C_{MD}$ represents more than just a measure of information. Intuitively, a state where $C_{MD} = 0$ is analogous to a coin-toss scenario in which coins A and B are completely independent, observing the result of coin B provides no information about coin A. In the context of quantum experiments, stating that $C_{MD} = 0$ implies that even with complete knowledge of the hidden variables, one cannot infer anything about the measurement settings.

In the model proposed by \citet{brans1988bell}, the settings of the measurement devices are fully determined by hidden variables, resulting in $C_{MD}=1$ \citep[p. 196]{hall2016significance}. Because measurement independence is completely violated in this case, the CHSH inequality is accordingly violated.

\citet[p. 197]{hall2016significance} demonstrated that by more cleverly adjusting the distribution of hidden variables $\lambda$, it is possible to significantly reduce the value of $C_{MD}$ while still violating the CHSH inequality. Surprisingly, even a minimal degree of correlation, equivalent to just $0.0663$ bits of mutual information between hidden variables and measurement settings \citep[p. 197]{hall2016significance}, is sufficient to violate the CHSH inequality. This suggests that full determinism is not required, challenging the necessity of strong superdeterminism. This mirrors the earlier coin example, where $0.0663$ bits of mutual information suggests only partial predictability between coin A and B. 

This finding challenges superdeterminism. Full determination of measurement settings by hidden variables is not required; even partial influence can suffice to violate the CHSH inequality. In this sense, superdeterminism can be regarded as an overly strong or extreme assumption.

As \citet{hall2016significance} points out, quantifying the degree of measurement independence is important not only from a philosophical standpoint but also from the perspective of quantum information theory. For example, \citet{ekert1991quantum} proposed quantum cryptographic protocols based on the violation of Bell inequalities. This proposal was qualitative in nature, but \citet{acin2007device} provided a rigorous information-theoretic proof of the security of quantum key distribution. \citet[p. 230501-1]{acin2007device} states that ``both Alice and Bob should be allowed to choose freely among at least two measurement settings,'' which implies an assumption of measurement independence.

If an eavesdropper possessed such a description, they could potentially determine the encryption key \citep[p. 190]{hall2016significance}. In the context of quantum key distribution, hidden variables are interpreted not as the initial conditions at the Big Bang, but rather as information potentially accessible to an eavesdropper. In this way, the philosophical issue of whether superdeterminism is valid is linked to the engineering question of whether quantum cryptographic protocols are secure.

\subsection{Classification of Intermediate Positions}
\label{classification-subsection}

As discussed above, Intermediate Positions 1, 2, and 3 can be understood as modifications of the positive heuristic of superdeterminism.



\subsubsection{First Intermediate Position: Soft Superdeterminism}
\label{first-intermediata-subsection}

In Intermediate Position 1 (Soft Determinism), the hidden variables correspond to the physical state of the detector. Its positive heuristic is that the measurement settings are determined by the physical states of the detectors. Since $C_{MD}$ represents the relationship between the detector's physical state and the measurement setting, it can be interpreted ontologically. Given that this position is deterministic, we have $C_{MD}=1$.

The toy model proposed by \citet{donadi2022toy} falls into the category of soft superdeterminism. They suggest that measurement outcomes may differ depending on whether the distribution of hidden variables is uniform or non-uniform. They also propose an experiment involving successive measurements of non-commuting observables using the same detector. Similarly, \citet{hossenfelder2011testing} propose an experiment to detect physically observable correlations between the measurement device and the prepared state of the system being measured. Since their aim is not to detect correlations with hidden variables dating back to the Big Bang, their proposal can also be interpreted as an attempt to test soft superdeterminism. Therefore, Intermediate Position 1 (Soft Determinism) may be experimentally testable.

\subsubsection{Second Intermediate Position: Partial Measurement Dependence Models}
\label{second-intermediata-subsection}

Intermediate Position 2 (Eavesdropper Model) interprets hidden variables as information potentially accessible to an eavesdropper in quantum cryptography. Its positive heuristic is that the eavesdropper knows the settings of the measurement devices. This allows for a partial violation of measurement independence ($C_{MD} > 0$), avoiding full determinism. The advantage of this position lies in its operational relevance to information security. 

When $C_{MD}=0$, the eavesdropper has no knowledge of the settings; when $C_{MD}=1$, the eavesdropper has complete knowledge. As discussed in Section \ref{correlation-subsubsection}, it is remarkable that an eavesdropper can violate the CHSH inequality with as little as $C_{MD} =0.0663$ bits of information.

From the standpoint of this intermediate view, a range of theoretical questions becomes accessible for investigation. For example, one can examine how much information an eavesdropper would need to compromise the security of quantum cryptographic protocols. One can also explore how such security might be empirically assessed. In this respect, the programme exhibits the potential to qualify as progressive.

\subsubsection{Third Intermediate Position: Application to Contextuality}
\label{third-intermediata-subsection}

Intermediate Position 3 (Adversarial RNG Model) is nearly identical to Intermediate Position 2 (Eavesdropper Model). The key difference lies in the framework: instead of the CHSH inequality, it is considered within the context of the KCBS inequality, which is related to noncontextuality. It is the idea that measurement outcomes are determined by intrinsic properties, independent of other measurements that could be performed simultaneously.

The foundational challenge posed by contextuality was first established in the Kochen-Specker theorem \citep{kochen1967problem}. This theorem demonstrates that, under certain assumptions, no single set of pre-assigned measurement values can remain consistent across all possible measurement contexts in quantum theory. Recent systematic reviews, such as \citet{budroni2022kochen}, highlight the increasing relevance of KCBS-type inequalities in this domain.

For a long time, Kochen-Specker violations remained theoretically important but experimentally elusive. That changed with the introduction of inequalities such as the Klyachko-Can-Binicio{\u{g}}lu-Shumovsky (KCBS) inequality, which made it possible to empirically test noncontextuality \citep{klyachko2008simple}. Unlike the CHSH inequality, which involves measuring two spatially separated systems, the KCBS inequality is tested on a single quantum system by examining a particular arrangement of measurements. Experiments have shown that quantum systems do indeed violate the KCBS inequality \citep{ahrens2013two}.

In conventional random number generation, pseudo-random behavior may arise from external environments or device noise. Therefore, it is essential to develop a method that guarantees, according to physical laws, that the randomness originates from the intrinsic unpredictability of quantum mechanics. To address this issue, a method has been proposed that uses the violation of the KCBS inequality to experimentally certify that the generated randomness stems from quantum contextuality \citep{um2013experimental, um2018correction}. This mirrors the role of CHSH inequality violations in quantum key distribution, where they serve to certify the absence of eavesdropping (Section \ref{second-intermediata-subsection}).

\citet{klyachko2008simple} derived the KCBS inequality without relying on assumptions such as outcome independence, parameter independence, and measurement independence. However, it has been shown that the KCBS inequality can also be derived from these assumptions \citep[Section 2]{kitajima2025states}. In this context, outcome independence and parameter independence are interpreted not as conditions related to locality but rather to noncontextuality. 

This opens the possibility of constructing a research programme analogous to the one discussed in Section \ref{second-intermediata-subsection}. Here, hidden variables are not considered as information potentially accessible to an eavesdropper, but instead as information held by a forger attempting to manipulate the measurement settings to forge random numbers. Its positive heuristic is that the forger knows the settings of the measurement devices. As a measure of the forger's knowledge of the measurement settings, $C_{MD}$ is best understood epistemologically. When $C_{MD}=0$, the forger has no knowledge of the settings; when $C_{MD}=1$, the forger has complete knowledge. 

While individual quantum states can violate the CHSH inequality and the KCBS inequality separately, it was believed that no single state could violate both simultaneously. This relationship suggests a trade-off between nonlocality and contextuality: if one is violated, the other cannot be. Such a relationship is known as a monogamy relation \citep{kurzynski2014fundamental}. 

\citet[Theorem 1]{xue2023synchronous} and \citet[Theorem 1]{kitajima2025states} demonstrated, however, that simultaneous violations of both the CHSH and KCBS inequalities are possible with specific choices of observables. One could also consider the question of how large $C_{MD}$ must be for both the CHSH inequality and the KCBS inequality to be violated simultaneously in a given state. If this question can be theoretically clarified, such a development could be regarded as progressive in the Lakatosian sense.

\subsection{Ontological and Epistemological Interpretations of Correlation Measure of Dependence $C_{MD}$}
\label{ontological-epistemological-subsection}

Correlation Measure of Dependence $C_{MD}$ between hidden variables and measurement settings can be interpreted in two distinct ways:

\begin{description}
\item[Ontological Interpretation:]
Correlation Measure of Dependence $C_{MD}$ reflects an objective feature of reality. It implies that measurement settings are causally influenced by hidden variables, possibly predetermined by the universe's initial conditions. In this view, the correlation is built into the fabric of the world and exists independently of our knowledge or ignorance.
\item[Epistemological Interpretation:] 
A nonzero Correlation Measure of Dependence $C_{MD}$ indicates limited knowledge or information leakage. It does not require a metaphysical commitment to determinism or hidden causal mechanisms. Instead, it reflects our imperfect control over experimental settings, such as the possibility that an adversary (e.g., an eavesdropper) might partially know or influence the measurement choices.
\end{description}


The interpretation of a nonzero $C_{MD}$ depends crucially on whether it is taken epistemologically or ontologically.

From an epistemological perspective, $C_{MD} > 0$ suggests limitations in our knowledge. This epistemological view holds that the observed dependence between hidden variables and measurement settings is not due to a fundamental causal link. Rather, it results from practical limitations in keeping measurement settings isolated from potential informational leakage.

In contrast, from an ontological perspective, $C_{MD} > 0$ implies the existence of non-trivial correlations in the causal structure of the universe. That is, measurement settings are not merely epistemically contaminated but are causally determined by hidden variables, potentially rooted in the universe's initial conditions.

Thus, the metaphysical weight carried by $C_{MD} > 0$ differs substantially depending on the interpretive stance. Clarifying which of these two views is adopted has significant implications for how one evaluates the validity and implications of intermediate positions.




When $C_{MD}$ is interpreted epistemologically, as in Intermediate Positions 2 and 3, it reflects our limited access to or control over measurement settings, rather than an objective feature of the world. This interpretation brings both advantages and drawbacks. The main drawback, from a metaphysical standpoint, is that it does not offer a clear ontological picture of how the universe ``truly is,'' in contrast to strong superdeterministic views. 

This apparent weakness, however, can also be viewed as a virtue of epistemic modesty. Non-relativistic quantum theory can account for violations of the CHSH inequality but cannot accommodate a description of the Big Bang. Therefore, the claims of superdeterminism can be seen as too strong from a theoretical standpoint. 

In contrast, Intermediate Position 2 (Eavesdropper Model) and Intermediate Position 3 (Adversarial RNG Model) avoid such strong claims by treating $C_{MD}$ as information possessed by an eavesdropper or someone attempting to forge randomness. By restricting the scope to problems that can be addressed within the framework of information theory, these positions exhibit epistemic modesty.





\subsection{Methodological Significance and Future Prospects of the Intermediate Position}
\label{future-prospects-subsection}


The points discussed in Sections \ref{classification-subsection} and \ref{ontological-epistemological-subsection} can be summarized in Table \ref{tab:comparison9}.

\begin{table}[htbp]
    \centering
    \renewcommand{\arraystretch}{1.3}
    \small
    \begin{tabular}{ccc}
        \toprule
        \makecell[c]{\textbf{Superdeterminism}\\\textbf{Intermediate Position 1}} &
        \makecell[c]{\textbf{Intermediate Position 2}\\\textbf{Intermediate Position 3}} &
        \textbf{Nonfactorizable Position} \\
        \midrule
        $C_{MD} = 1$ & $0 \leq C_{MD} \leq 1$ & $C_{MD} = 0$ \\
        Deterministic & Probabilistic correlation & Complete independence \\
        \midrule
        \makecell[c]{$C_{MD}$ reflects\\an ontological situation}  & \makecell[c]{$C_{MD}$ reflects\\an epistemological situation} & \makecell[c]{$C_{MD}$ reflects\\an epistemological situation} \\
        \bottomrule
    \end{tabular}
    \caption{Spectrum of the Intermediate Positions}
    \label{tab:comparison9}
\end{table}

Particularly notable are the approaches developed by \citet{hall2011relaxed, hall2016significance} and \citet{barrett2011much}, which employ information-theoretic measures, such as mutual information, to quantify the extent to which measurement independence is violated. Within this framework, the dependence between measurement settings and hidden variables is modeled as a continuous parameter $C_{MD}$ allowing for the formal treatment of scenarios where measurement independence only partially holds.

Measurement independence is fundamentally an epistemological condition. The nonfactorizable position assumes $C_{MD}=0$. From this, it is natural to explore scenarios with $C_{MD}>0$, such as the Eavesdropper Model in Intermediate Position 2. This position, like the nonfactorizable position, interprets $C_{MD}$ as an epistemological indicator of the information available to an external agent.

Intermediate Position 2 (Eavesdropper Model) offers at least three theoretical advantages over full-fledged superdeterminism. First, they retain sufficient explanatory power to account for violations of the CHSH inequality without invoking complete determinism, in which settings are fixed by hidden variables originating at the Big Bang. Second, the use of quantifiable indicators like $C_{MD}$ creates a natural link to practical applications in quantum cryptography and information security, thereby bridging philosophical interpretations and engineering implementations. Third, as discussed in Section \ref{third-intermediata-subsection}, the concept of $C_{MD}$ can potentially be applied not only within the framework of the CHSH inequality but also within the framework of the KCBS inequality. This is Intermediate Position 3 (Adversarial RNG Model).

Furthermore, Intermediate Positions 2 and 3 may be considered intelligible in de Regt's sense. In these positions, $C_{MD}$ is interpreted as information held by an eavesdropper or a forger. Under this interpretation, one can qualitatively predict that as the amount of such information increases, the level of security decreases. This aligns with de Regt's notion of intelligibility. If this qualitative prediction is theoretically demonstrated and empirically confirmed, the position could become fruitful in Kuhn's sense and progressive in Lakatos's sense.

\section{Conclusion}

This paper has examined two competing responses to the violation of the CHSH inequality: the nonfactorizable position and superdeterminism. These perspectives were evaluated using three major frameworks from the philosophy of science:
\begin{enumerate}
\item de Regt's account of scientific understanding,
\item Kuhn's criteria for theory choice, and
\item Lakatos's theory of research programmes.
\end{enumerate}

While all three frameworks, de Regt's, Kuhn's, and Lakatos's, address theory evaluation, they do so from philosophically distinct angles. De Regt focuses on the intelligibility of theories for practitioners, Kuhn provides multifaceted criteria for theory choice, and Lakatos evaluates the long-term progressiveness of research programmes. Their integration allows for a multidimensional assessment: how theories are understood (de Regt), chosen (Kuhn), and developed (Lakatos). These frameworks are therefore complementary, not redundant.

The findings are as follows:

\begin{enumerate}
\item According to de Regt's Criterion of Intelligibility, the nonfactorizable position offers a more accessible, qualitative understanding of quantum phenomena, as demonstrated by quantum teleportation. Superdeterminism, by contrast, lacks intelligibility (Section \ref{de Regt-section}).
\item While Kuhn's framework allows for subjective weighting of theory choice criteria, emphasizing consistency and fruitfulness leads to a preference for the nonfactorizable position (Section \ref{kuhn-comparison-subsection}).
\item From Lakatos's perspective, the nonfactorizable position constitutes a progressive research programme, producing novel predictions and empirical success. While superdeterminism currently appears to be a degenerative research programme due to its limited empirical applications, future developments may revise this assessment. (Section \ref{Lakatos-section}).
\end{enumerate}

These evaluations are summarized in Table \ref{tab:comparison2}. This paper has assessed the two positions through a multi-framework philosophical analysis. When combined with the conclusions drawn from the frameworks of de Regt, Kuhn, and Lakatos, it becomes evident that the nonfactorizable position is the more promising of the two\footnote{Recent results have shown that quantum nonlocality can emerge when the settings of the measuring devices are not fully determined, that is, when the probabilities of those settings, given the hidden variables, are neither $0$ nor $1$. \citep{putz2014arbitrarily, supic2023quantum}. This further supports the robustness of the nonfactorizable position.}.

\begin{table}[htbp]
    \centering
    \renewcommand{\arraystretch}{1.3}
    \small
    \begin{tabular}{lcc}
        \toprule
        & \textbf{\large Nonfactorizable} & \textbf{\large Superdeterminism} \\
        & \textbf{\large Position} & \\
        \midrule
        \textit{de Regt} & Intelligible & Not intelligible \\
        \midrule
        \textit{Kuhn} & \makecell[c]{Consistent\\Fruitful} & \makecell[c]{Not consistent\\Not fruitful} \\
        \midrule
        \textit{Lakatos} & Progressive & Degenerative \\
        \bottomrule
    \end{tabular}
    \caption{Comparison Between the Nonfactorizable Position and Superdeterminism}
    \label{tab:comparison2}
\end{table}


In Section \ref{intermediate-section}, we examined intermediate positions situated between superdeterminism and the factorizable position. As shown in Table \ref{tab:comparison7.1}, these positions represent revisions to the positive heuristic of superdeterminism. In the course of these revisions, we introduced a quantity, $C_{MD}$, to quantify measurement dependence. In Section \ref{future-prospects-subsection}, we argued that Intermediate Positions 2 and 3, where $C_{MD}$ is interpreted as the amount of information possessed by an eavesdropper or a forger, have the potential to become progressive research programmes in the Lakatosian sense.

Our analysis not only reframes existing debates about Bell's inequality but also illustrates how philosophical tools can deepen our understanding of theory appraisal in physics. While the discussion of $C_{MD}$ and cryptographic security may appear engineering-oriented, such practical concerns have historically driven major theoretical advances, as exemplified by the origins of quantum theory in blackbody radiation. In this way, the study not only reframes existing debates on quantum foundations, but also demonstrates how methodological tools from philosophy of science can actively shape emerging research agendas.

\section*{Acknowledgments}
We would like to thank the reviewers for their valuable comments. 


\bibliographystyle{sn-chicago}

\end{document}